\documentclass[referee]{RAA}            
\usepackage{graphicx,times}             
\usepackage[]{natbib}
\input{epsf.sty}                        
\input{psfig.sty}                       

\begin{document}

   \title{New Supernova Candidates from SDSS-DR7 of Spectral Survey
\,$^*$ \footnotetext{$*$ Supported by the National Natural Science
Foundation of China.} }

   \volnopage{Vol.0 (200x) No.0, 000--000}      
   \setcounter{page}{1}          

   \author{Liang-Ping Tu
      \inst{1,2,3,\star}
   \and A-Li Luo
      \inst{2,\star}
   \and Fu-Chao Wu
      \inst{1}
   \and Chao Wu
      \inst{2}
   \and Yong-Heng Zhao
      \inst{2}
   }


   \institute{Institute of Automation, Chinese Academy of Sciences,
             Beijing 100190, China; \\{\it lptu@nlpr.ia.ac.cn}\\
        \and
             National Astronomical Observatories,Chinese Academy of Sciences,
             Beijing 100012, China; {\it lal@bao.ac.cn}\\
        \and
             School of Science, University of Science and Technology Liaoning,
             Anshan 114051, China\\
   }

   \date{Received~~2009 month day; accepted~~2009~~month day}

\abstract{The letter presents 25 discovered supernova candidates
from SDSS-DR7 with our dedicated method, called Sample Decrease, and
10 of them were confirmed by other research groups, and listed in
this letter. Another 15 are first discovered including 14 type Ia
and one type II based on Supernova Identification (SNID) analysis.
The results proved that our method is reliable, and the description
of the method and some detailed spectra analysis procedures were
also presented in this letter.} 
   \authorrunning{Liang-Ping Tu,A-Li Luo,Fu-Chao Wu,Chao Wu \& Yong-Heng Zhao }            
   \titlerunning{New Supernova Candidates from SDSS-DR7 of Spectral Survey }  

   \maketitle

%
%
\section{Introduction}           
\label{sect:intro}

Supernovae (SNe) are generally discovered by repeat imaging of the
same region of sky every other night, and measuring light curves for
objects in the area. The SDSS Supernova Survey was one of three
surveys (along with the Legacy and SEGUE surveys) of SDSS-II, a
3-year extension of the original SDSS that operated from July 2005
to July 2008. Besides the imaging survey, the spectral survey of
SDSS also gathers large amount of spectroscopy of galaxies. These
spectra are basis of many astronomical research. As a byproduct, SNe
in their host galaxies are possible to be detected spectroscopically
Madgwick et al.~\cite{madg03}, since a SN spectrum might have
obvious broad peaks and troughs that modulate the spectrum of its
host galaxy. By the rate of SNe detection computed in Madgwick et
al.~\cite{madg03}, there are $\sim200$ Type Ia SNe detection of
$\sim10^{6}$ galaxy spectra in SDSS-DR7(Abazajian et al. 2008).
Madgwick et al.~\cite{blon07} have detected 19 type Ia SNe from
$\sim10^{5}$ galaxy spectra in SDSS-DR1(Abazajian et al. 2003)
through the spectroscopic approach.\\
In this Letter, we reported the 14 Type Ia SNe and 1 Type II SN
which are detected through spectroscopic approach in SDSS-DR7. Our
method is some different with Madgwick et al. (2003), and is
described in detail in \S2, the detecting results are described in
\S3, and conclusion and future possibilities are summarized in \S4.


\section{Method}
\label{sect:method}

The spectroscopic approach mentioned by Madgwick et al.
~\cite{madg03} needs to perform galaxy subtraction and match with
all templates for each spectrum. To simplify the procedure in a
large number of spectra dataset, we present a concept of \lq\lq
Sample Decrease\rq\rq, namely before confirming SNe by template
matching, the most of galaxy spectra in the dataset without obvious
SN features are excluded, only those possible candidates are kept to
complete the template matching.

\subsection{Sample decrease}

The host-galaxy spectra with SN are much more sparse than galaxy
spectra without SN, so we can remove the most of them using outlier
detection method. The process could be divided into two phases, the
first phase is SN statistic eigen representation of each galaxy
spectrum, and the second phase is outlier detection. Here, Ia-normal
templates in the Peter Nugents' SN Spectral Templates library(Nugent
P. 1997) are used as our templates, which include 6th$\sim$40th
templates of all 90 templates. The reasons of such template
selection are that the SN characters of them are obvious and easy to
detect. Linear interpolation are performed to each of these 35 SN
templates to wavelength range 3801$\sim$8000${\AA}$ firstly, then
these 35 spectra are transformed by Principal Component Analysis
(PCA) to obtain the 12 eigen spectra (eigenvectors), which span an
eigen-subspace of Type Ia SN. The projection of one normalized
galaxy spectrum on these space is a 12-dimension vector, and it is
called a SN statistic eigen-representation of this galaxy spectrum.
The first two dimension eigenspectrum (Principal component) are
shown in
Figure 1.\\
   \begin{figure}
   \centering
   \includegraphics[width=\textwidth, height=3.5cm, angle=0]{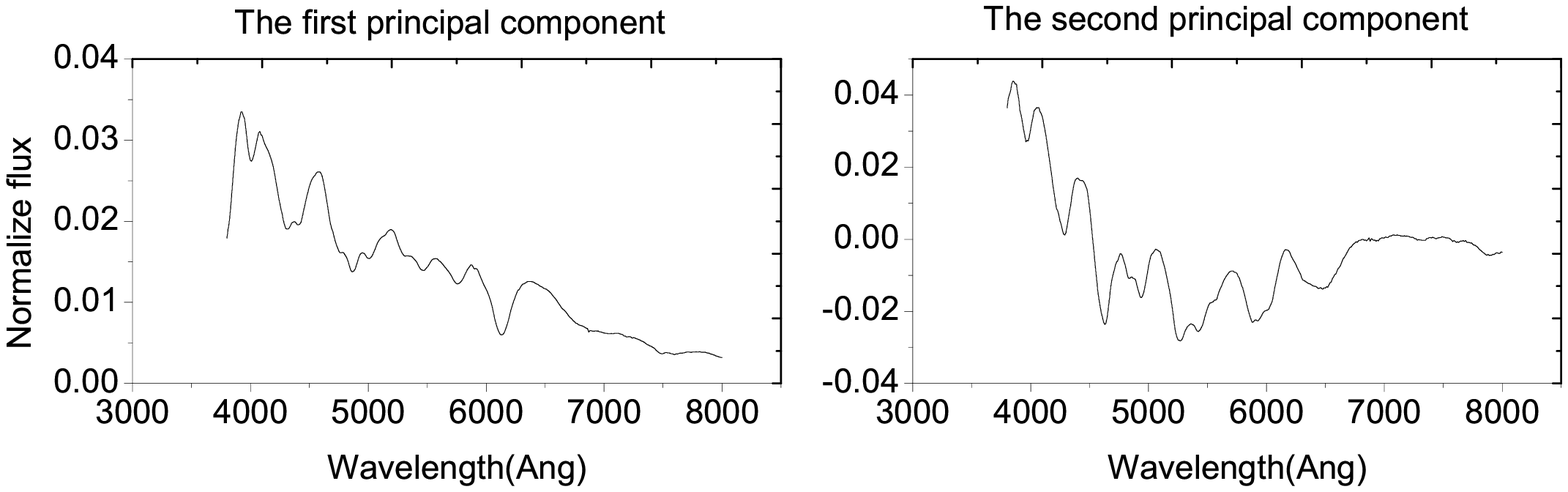}
   \vspace{-1cm}
   \caption{The First and Second Principal Components of Ia Supernova Templates}
   \label{Fig:Pca}
   \end{figure}
Markus M. ~\cite{mark00} has presented the conception of Local
Outlier Factor(LOF) used for outlier detection, which can be applied
to describe the singularity of galaxy spectra with SN component in
all galaxy spectra. Thus, the outlier might be spectra of SNe plus
their host galaxies. The
definitions related to LOF are described as follows:\\
\textbf{Definition 1:} ($k-distance$ of an object $p$). For any
positive integer $k$, the $k$-distance of object $p$, denoted as
$k-distance(p)$, is defined as the distance $d(p,o)$ between $p$ and
an object $o\in D$ such that: \\(i) For at least $k$ objects $o'\in
D\setminus p$, it holds that $d(p,o')\leq d(p,o)$\\  (ii) For at
most $k$-1 objects $o'\in D\setminus p$,
it holds that $d(p,o')<d(p,o)$.\\
\textbf{Definition 2:}($k-distance$ neighborhood of an object $p$).
Given the $k-distance$ of $p$, the $k-distance$ neighborhood of $p$
contains every object whose distance from $p$ is not greater than
the $k-distance$, i.e.$N_{k-distanc(p)}(p)=\{q\in D\setminus p\mid
d(p,q)\leq k-distance(p)\}$. These objects $q$ are called the
$k-nearest$ neighbors of $p$. Simplify the notation to use
$N_{k}(p)$ as a shorthand for $N_{k-distance}(p)$.\\
\textbf{Definition 3:} (reachability distance of an object $p$
$w.r.t$ object $o$). Let $k \in Z^{+}$, the reachability distance of
object $p$ with respect to object $o$ is defined as
$reachdist_{k}(p,o)=max\{k-distance(o),d(p,o)\}$.\\
\textbf{Definition 4 \& 5:} The local reachability density and the
local outlier factor of an
object $p$ are defined as\\
\begin{equation}
lrd_{k}(p)=1/\frac{\sum_{o\in
N_{k}(p)}reachdist_{k}(p,o)}{|N_{k}(p)|}
\end{equation}
\begin{equation}
LOF_{k}(p)=\frac{\sum_{o\in
N_{k}(p)}\frac{lrd_{k}(o)}{lrd_{k}(p)}}{|N_{k}(p)|}
\end{equation}

Using the above \emph{k}-LOF definition, we can design the
procedures of data reduction:\\
(i)   calculate the $k-distance$ of each sample in \emph{D};\\
(ii)  ascertain the $k-distance$ neighborhood of each object $p\in
D$;\\
(iii) calculate the local reachability density of each object $p\in
D$ using formula (1);\\
(iv)  calculate the k-local outlier factor of each object $p\in D$
using formula (2);\\
(v)   array all of the objects in \emph{D} according to their
\emph{k}-LOF in descending order, and keep the objects which have
higher \emph{k}-LOF as the range of candidates, here we keep
$1\sim2$ percent of all samples.

\subsection{Cross-correlation template matching based on SNID}

After sample decrease in SDSS-DR7, the number of remainder is 2945.
Template matching should be completed for this relative small
dataset. Blondin et al.~\cite{blon07} presented an interactive
cross-correlation method named SNID to identify SNe, and get the
redshift, age, and type of each SN, since all template spectra are
at zero redshift, and type and age of each template is known. We
simplify the SNID procedure into first four steps as follow and make
it running automatically, then the 5th step is to run SNID pipeline
to
confirm selected samples with high probability to be SN: \\
(i) Bin each spectrum on a logarithmic wavelength axis. \\
(ii) Continuum removal and spectra normalizing. \\
(iii) Smooth the spectra using mean filter, remove strong lines in
each spectrum.\\
(iv) Matching by cross-correlation, details can be found in
Blondin et al.~\cite{blon07} and Tonry et al.~\cite{tonry79}.\\
(v) Each candidate with high
confidence as a SN should be checked interactively through SNID pipeline. \\
An example of a spectrum which is addressed according to the former
steps is shown in Figure 2.

   \begin{figure}
   \centering
   \includegraphics[width=\textwidth, height=8cm, angle=0]{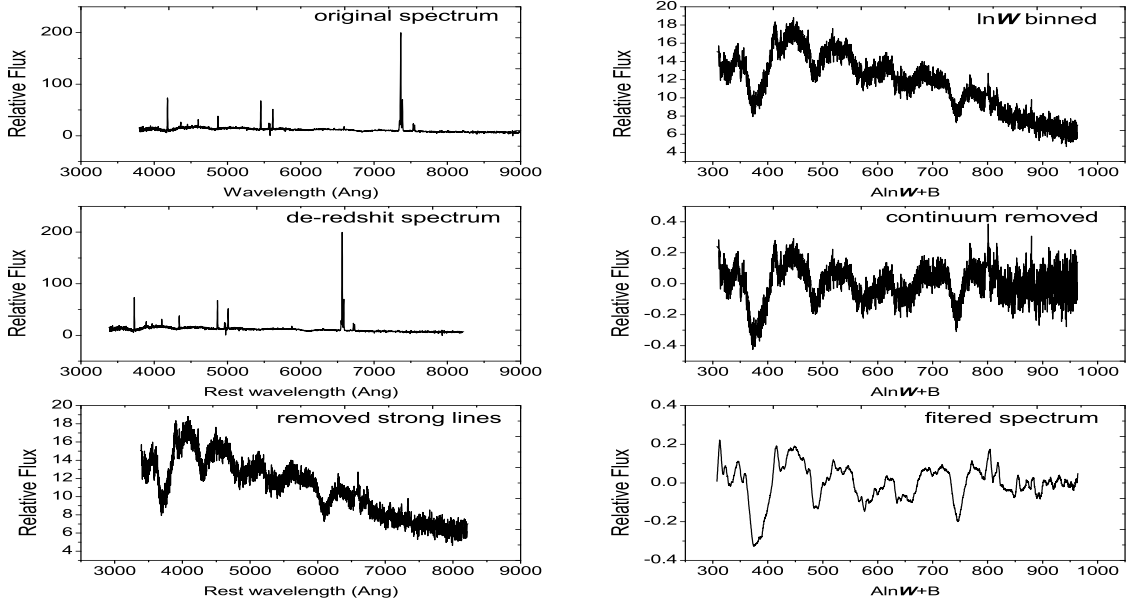}
   \caption{An example of SN spectrum. The top left shows original spectrum,
   the middle left is the corresponding spectrum which has been de-redshift,
   and the left bottom shows the result that strong lines are removed.The binned
   spectrum, the continuum divided spectrum and the smoothed spectrum are
   shown in the right three panels respectively. }
   \label{Fig:precess}
   \end{figure}

\section{Results}
\label{sect:Res}

We applied our method to all fields of SDSS DR7 (survey plate number
0266 $\sim$ 2974), excluding the spectra with low signal to noise
ratio (S/N) and with uncertainty of high redshift. The selection
criteria are: S/N$>$10 (all in g,r,i band), z$<$0.25, and with the
spectral type \lq\lq galaxy\rq\rq. The total number of galaxy
spectra is 294843, in which we performed the sample decrease
procedure. First step is to de-redshift for all galaxy spectra, like
figure 2 shows. Secondly, strong lines were removed in each spectrum
using narrow scaled wavelet filter, just to keep broad peaks and
troughs such as SiII absorption at 6150\AA, which is one of main
characters of Ia SN spectrum. A spectrum that strong lines are
removed is shown in the left bottom panel in Figure 2. After
pre-processing, we calculated SN statistical representations for all
spectra, i.e. projected each normalized spectrum on the
eigen-subspace. Then, we compute and sort the LOF of each sample,
and prune 99\% samples with low LOFs in DR7 galaxy dataset. Finally,
we obtained a total of 2945 spectra from the
 all 294843 spectra.\\
We then used cross-correlation template matching technique (Tonry et
al. 1979; Blondin et al. 2007) to calculate the similarities between
a spectrum and each template, and sort them in descending order. By
comparing all similarities in all template, only 65 spectra were
left to be identified. Finally, we confirmed 25 spectra of them as
SN through SNID. Among these 25 spectra,  5 of them have been
discovered by Madgwick et al.~\cite{madg03}(see top 5 rows in table
1), the other 5 spectra have been recorded elsewhere (see bottom 5
rows in table 1). The remainder of them have not been recorded,
moreover, the 15 spectra are shown in Figure 3 and the parameters of
them are listed in table 2.

%

\begin{table}
\begin{center}
\caption[]{The recorded SN+host spectra that we discovered.}
\label{Tab:publ-works}


 \begin{tabular}{lllllll}
  \hline

  SDSS Name             &IAU name   &Date        & $z$     &Age(days)       &Type   &SDSS
  r-mag\\
  \hline
  SDSS J101800.47-000157.9      &Sn 2000fx  &2000-12-01    & 0.065  &16-18     &Ia-pec   &17.98\\
  SDSS J080312.61+473649.7      &Sn 2000fy  &2000-12-06    & 0.117  &2.5-3.5   &Ia-norm  &17.93\\
  SDSS J011835.83+144100.5      &Sn 2000fz  &2000-12-15    & 0.054  &6.3-6.8   &Ia-norm  &16.73\\
  SDSS J092229.14+575429.3      &Sn 2001kj  &2001-01-02    & 0.063  &10-11     &Ia-norm  &17.79\\
  SDSS J095153.07+010605.7      &Sn 2001kp  &2001-03-21    & 0.063  &-3 to -2  &Ia-pec   &17.58\\
  SDSS J095915.75+005802.3      &Sn 2001kr  &2001-03-26    & 0.086  & 9.4      &Ia-csm   &18.15\\
  SDSS J093749.92+101138.0      &Sn 2003ly  &2003-12-17    & 0.095  & 17.00    &Ia-norm  &18.46\\
  SDSS J095948.15+112825.3      &Sn 2004ar  &2004-02-20    & 0.064  & 9.80     &Ia-norm  &18.04\\
  SDSS J132834.01+415108.2      &Sn 2004co  &2004-04-17    & 0.029  & 8.1      &Ia-norm  &15.82\\
  SDSS J154024.75+325157.2      &Sn 2004cp  &2004-05-24    & 0.053  & 28.5     &Ia-norm  &17.50\\
  \hline

\end{tabular}
\end{center}
\end{table}

\begin{table}
\begin{center}
\caption[]{ The new discovered SN+host spectra.}
 \begin{tabular}{lllllll}
  \hline\noalign{\smallskip}
SDSS Name       &Another name   &Date    &$z$     &Age(days)       &Type   &SDSS  r-mag\\
\hline

SDSS J074734.48+272647.4   &anon                       &2002-12-10   &0.061  &41.6   &Ia-91bg    &17.45\\
SDSS J074933.17+275729.4   &anon                       &2002-12-10   &0.131  &-5     &Ia-norm    &18.42\\
SDSS J112900.54+484359.2   &anon                       &2003-01-03   &0.085  &-3.4   &Ia-norm    &18.01\\
SDSS J081647.02+251731.6   &anon                       &2003-03-11   &0.158  &-8.6   &Ia-norm    &18.11\\
SDSS J160132.55+265915.1   &anon                       &2003-07-02   &0.071  &-2     &Ia-norm    &16.64\\
SDSS J083909.66+072431.6   &2MASX  J08390967+0724320   &2003-11-21   &0.042  &23.7   &Ia-91T     &17.29\\
SDSS J113913.53+150215.7   &anon                       &2005-01-16   &0.022  &34.1   &IIp        &18.36\\
SDSS J104440.53+303803.3   &anon                       &2005-02-28   &0.077  &4.3    &Ia-norm    &18.09\\
SDSS J160116.52+174603.9   &MCG+03-41-035              &2005-07-05   &0.040  &16.1   &Ia-norm    &16.46\\
SDSS J162423.71+154036.2   &anon                       &2005-07-07   &0.097  &0.30   &Ia-91T     &17.85\\
SDSS J114438.44+295323.7   &anon                       &2006-03-03   &0.075  &19.1   &Ia-norm    &17.94\\
SDSS J084943.93+121755.3   &NGC 2682                   &2006-03-21   &0.053  &28.5   &Ia-norm    &18.39\\
SDSS J105710.63+092403.4   &anon                       &2008-02-02   &0.086  &9.4    &Ia-csm     &18.12\\
SDSS J132301.40+243023.6   &NGP9 F380-0370514          &2008-02-28   &0.088  &0.3    &Ia-norm    &17.43\\
SDSS J150531.71+175904.7   &2MASX J15053172+1759051    &2008-03-31   &0.033  &14.1   &Ia-norm    &16.23\\

\noalign{\smallskip}\hline
\end{tabular}
\end{center}
\end{table}

\section{Conclusion}
\label{sect:Con} A novel spectroscopic analysis of the $\sim300000$
galaxy spectra in the SDSS-DR7 has resulted in the definite
identification of 14 SNe Ia and one SN II. There are three reasons
for not ensuring search completeness. The first is that reduced
samples were fixed to 1\% of all samples to keep high reliability,
and some SNe would be lost. The second is that only half of Nugents'
SN templates with obvious SN characters were adapted in this method.
The last is that there is still a problem of host galaxy
subtraction, and we have to give up searching those faint SNe. In
this method, only small samples are matched with templates, which
improved search efficiency.

%

%

   \begin{figure}
   \centering
   \includegraphics[width=\textwidth, height=4.5cm, angle=0]{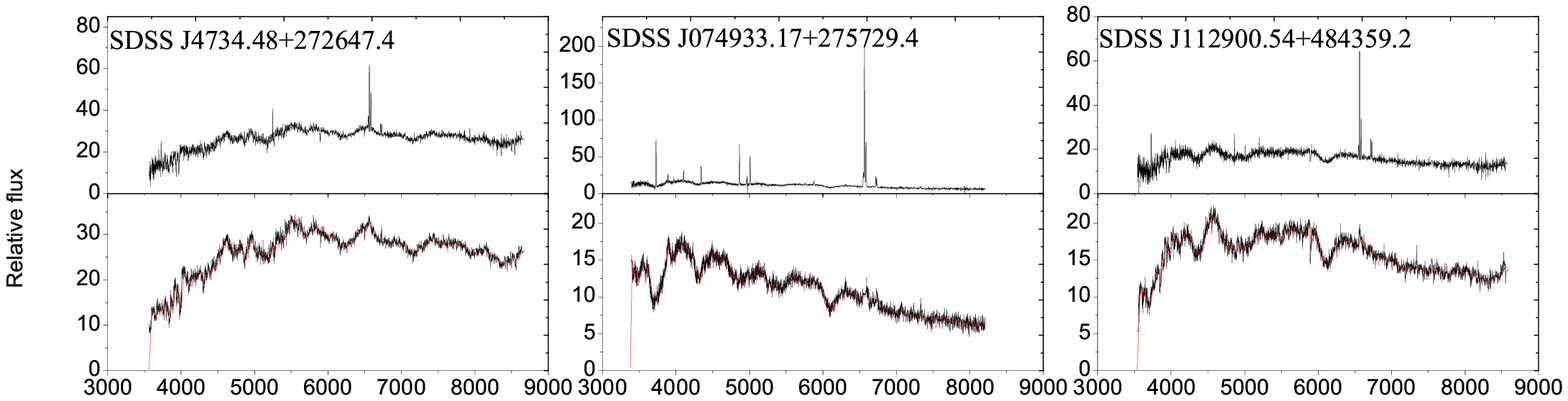}
   \vspace{-0.1cm}
   \includegraphics[width=\textwidth, height=4.5cm, angle=0]{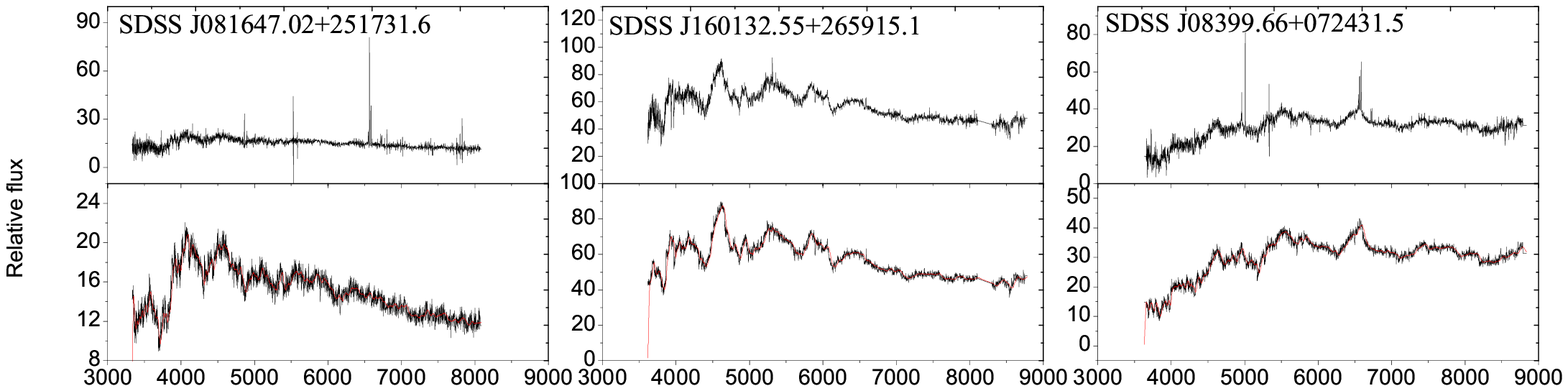}
   \vspace{-0.1cm}
   \includegraphics[width=\textwidth, height=4.5cm, angle=0]{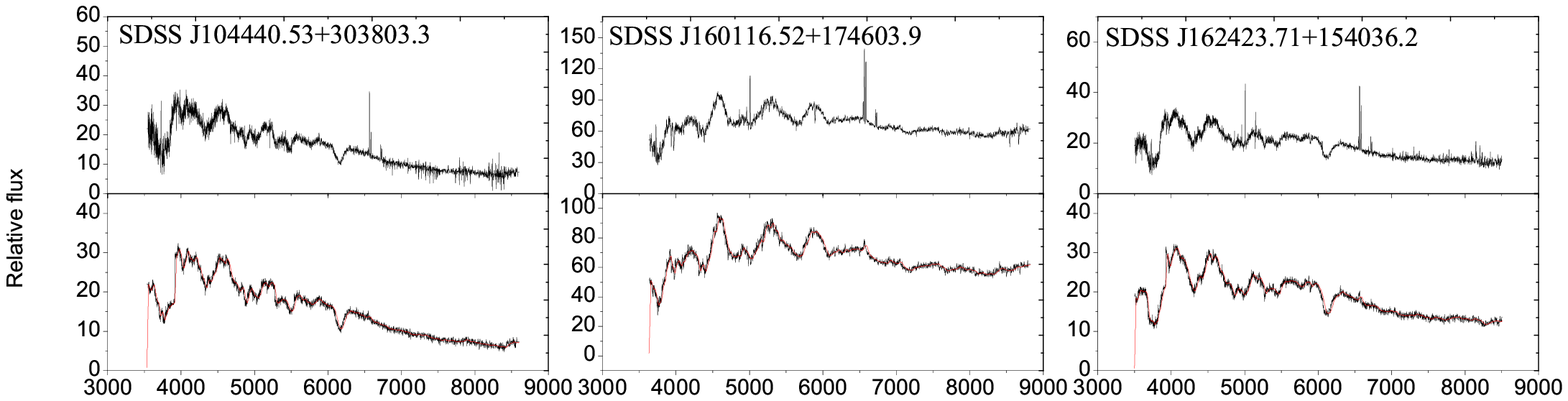}
   \vspace{-0.1cm}
   \includegraphics[width=\textwidth, height=4.5cm, angle=0]{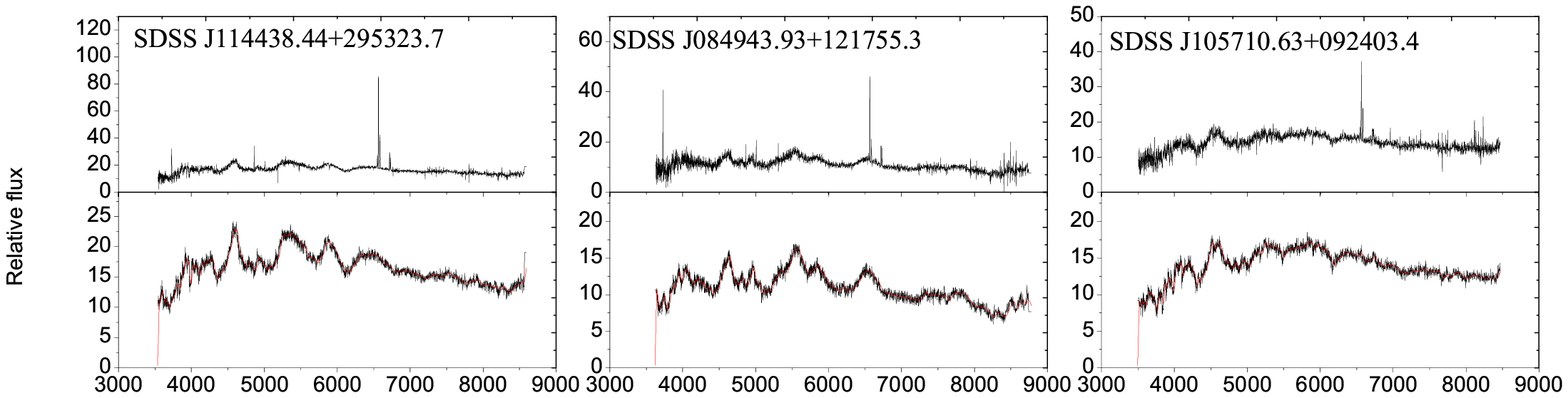}
   \vspace{-0.1cm}
   \includegraphics[width=\textwidth, height=4.8cm, angle=0]{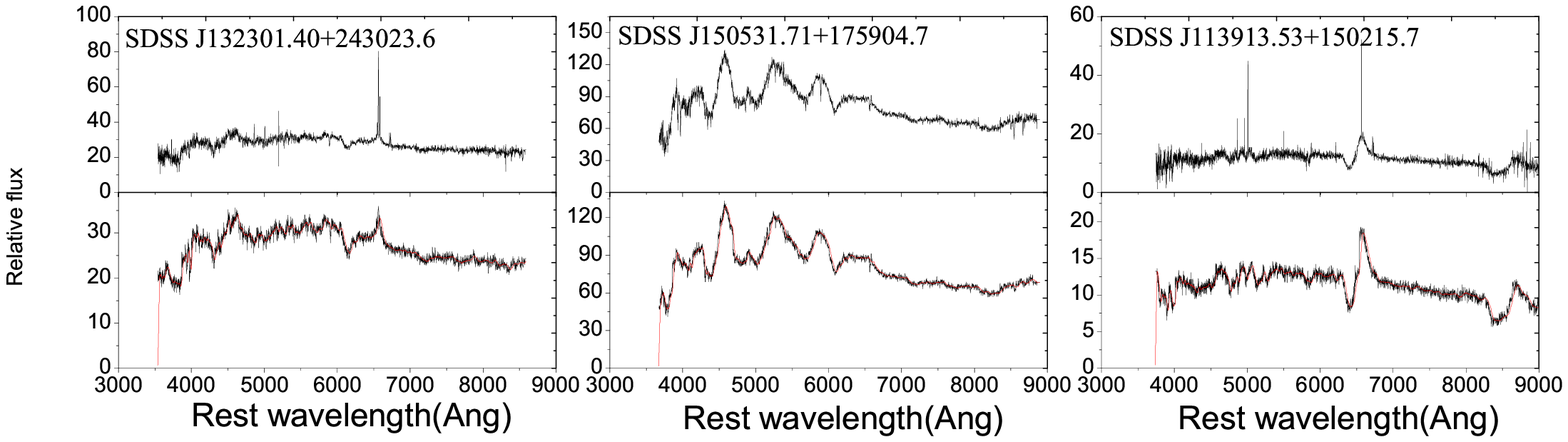}
   \caption{spectra of the new SNe identified in SDSS-DR7 are shown in
   each panel. In addition, the bottom plot in each panel shows the
   spectrum removed strong lines. The smooth spectrum is shown overplotted
   on this spectrum.}
   \label{Fig:sn1}
   \end{figure}

\begin{acknowledgements}
The authors are grateful to thank Jingyao HU, Jinsong Deng, Yulei
Qiu for many useful discussions and the referee for helpful
suggestions. All experiment are based on SDSS data releases. This
work were funded by the National Natural Science Foundation of China
(NSFC) under No.60773040 and No.60402041, and supported by the Young
Researcher Grant of NAOC, CAS.
\end{acknowledgements}

\label{lastpage}

\end{document}